\documentclass[journal]{IEEEtran}
\usepackage[pdftex]{graphicx}
\usepackage{url}

\ifCLASSINFOpdf
\else
\fi
\begin{document}
%
% paper title
% Titles are generally capitalized except for words such as a, an, and, as,
% at, but, by, for, in, nor, of, on, or, the, to and up, which are usually
% not capitalized unless they are the first or last word of the title.
% Linebreaks \\ can be used within to get better formatting as desired.
% Do not put math or special symbols in the title.
\title{Design and performance of the TIGER\\front-end ASIC for the BESIII Cylindrical\\Gas Electron Multiplier detector}
%
%
% author names and IEEE memberships
% note positions of commas and nonbreaking spaces ( ~ ) LaTeX will not break
% a structure at a ~ so this keeps an author's name from being broken across
% two lines.
% use \thanks{} to gain access to the first footnote area
% a separate \thanks must be used for each paragraph as LaTeX2e's \thanks
% was not built to handle multiple paragraphs
%

\author{
	Fabio~Cossio$^{*}$, Maxim Alexeev, Ricardo~Bugalho, Junying~Chai, Weishuai~Cheng, Manuel~D.~Da~Rocha~Rolo,~Agostino~Di~Francesco, Michela~Greco, Chongyang~Leng, Huaishen~Li, Marco~Maggiora, Simonetta~Marcello,~Marco~Mignone, Angelo~Rivetti, Joao~Varela and Richard~Wheadon%
	\thanks{$^{*}$ Corresponding author (e-mail: fcossio@to.infn.it).}% <-this % stops a space	
	\thanks{F. Cossio is with the Politecnico of Turin, Italy and also with the I.N.F.N., Section of Turin, Italy.}% <-this % stops a space
	\thanks{M. Alexeev, M. Greco, M. Maggiora and S. Marcello are with the Department of Physics, University of Turin, Italy and also with the I.N.F.N., Section of Turin, Italy.}% <-this % stops a space
	\thanks{R. Bugalho is with PETSys Electronics, Oeiras, Portugal.}% <-this % stops a space	
	\thanks{J. Chai, W. Cheng and C. Leng are with the Institute of High Energy Physics, Beijing, China, and also with the University of Chinese Academy of Sciences, China, and Politecnico of Turin, Italy and the I.N.F.N., Section of Turin, Italy.}% <-this % stops a space
	\thanks{M. D. Da Rocha Rolo, M. Mignone, A. Rivetti and R. Wheadon are with the I.N.F.N., Section of Turin, Italy.}% <-this % stops a space
	\thanks{A. Di Francesco is with LIP, Lisbon, Portugal.}% <-this % stops a space	
	\thanks{H. Li is with the Institute of High Energy Physics, Beijing, China, and also with the I.N.F.N., Section of Turin, Italy.}% <-this % stops a space
	\thanks{J. Varela is with PETSys Electronics, Oeiras, Portugal and LIP, Lisbon, Portugal.}% <-this % stops a space	
	\thanks{Manuscript received November 17, 2017.}% <-this % stops a space
}

\begin{figure*}[!t]
%\centering
{\Large This paper is a preprint (IEEE ``accepted'' status). It has been published in IEEE Xplore Proceedings for “2017 IEEE Nuclear Science Symposium and Medical Imaging Conference (NSS/MIC)”

\vspace{0.5cm}

DOI: \url{10.1109/NSSMIC.2017.8533025}
}

\vspace{1cm}
{\Large
\textbf{IEEE copyright notice}
}
\vspace{0.5cm}
{\large
	
\copyright 2019 IEEE. Personal use of this material is permitted. Permission from IEEE must be obtained for all other uses, in any current or future media, including reprinting/republishing this material for advertising or promotional purposes, creating new collective works, for resale or redistribution to servers or lists, or reuse of any copyrighted component of this work in other works.}

\vspace{12cm}
%\subfloat[Case I]{\includegraphics[width=2.5in]{box}%
%\label{fig_first_case}}
%\hfil
%\subfloat[Case II]{\includegraphics[width=2.5in]{box}%
%\label{fig_second_case}}
%\caption{Simulation results for the network.}
%\label{fig_sim}
\end{figure*}

% make the title area
\maketitle

% As a general rule, do not put math, special symbols or citations
% in the abstract or keywords.
\begin{abstract}
We present the design and characterization of TIGER (Turin Integrated Gem Electronics for Readout), a 64-channel ASIC developed for the readout of the CGEM (Cylindrical Gas Electron Multiplier) detector, the proposed inner tracker for the 2018 upgrade of the BESIII experiment, carried out at BEPCII in Beijing.

Each ASIC channel features a charge sensitive amplifier coupled to a dual-branch shaper stage, optimized for timing and charge measurement, followed by a mixed-mode back-end that extracts and digitizes the timestamp and charge of the input signals. The time-of-arrival is provided by a set of low-power TDCs, based on analogue interpolation techniques, while the charge measurement is obtained either from the Time-over-Threshold information or with a sample-and-hold circuit.

The ASIC has been fabricated in a 110 nm CMOS technology and designed to operate with a 1.2 V power supply, an input capacitance of about 100 pF, an input dynamic range between 3 and 50 fC, a power consumption of about 12 mW/channel and a sustained event rate of 60 kHz/channel. The design and test results of TIGER first prototype are presented showing its full functionality.
\end{abstract}

% Note that keywords are not normally used for peerreview papers.

% For peer review papers, you can put extra information on the cover
% page as needed:
% \ifCLASSOPTIONpeerreview
% \begin{center} \bfseries EDICS Category: 3-BBND \end{center}
% \fi
%
% For peerreview papers, this IEEEtran command inserts a page break and
% creates the second title. It will be ignored for other modes.
\IEEEpeerreviewmaketitle

\IEEEpubid{978-1-5386-2282-7/17/\$31.00 \copyright2017 IEEE}
\IEEEpubidadjcol

\section{Introduction}

\IEEEPARstart{B}{ESIII} (BEijing Spectrometer III) is a multipurpose detector to study the collisions provided by the Beijing Electron-Positron Collider II (BEPCII), hosted at the IHEP (Institute of High Energy Physics) laboratory~\cite{BESIII}. The Inner Tracker is showing aging effects due to the high radiation absorbed and since BESIII will run at least until 2022, it is foreseen to upgrade the Inner Tracker with a new detector. The proposed new inner tracker comprises three layers of Cylindrical Gas Electron Multiplier (CGEM) detector~\cite{sauli, GEMTDR}. Each layer consists of a cathode, three GEMs foils and the readout anode, which is segmented with 650 $\mu$m pitch XV-patterned strips. In order to achieve the required spacial resolution (130 $\mu$m), an analogue readout has been chosen since it allows to employ charge centroid and micro-TPC (Time Projection Chamber) algorithms to reduce the total number of channels to about 10.000~\cite{lia}.

\begin{figure}
	\centering
	\includegraphics[width=3.0in]{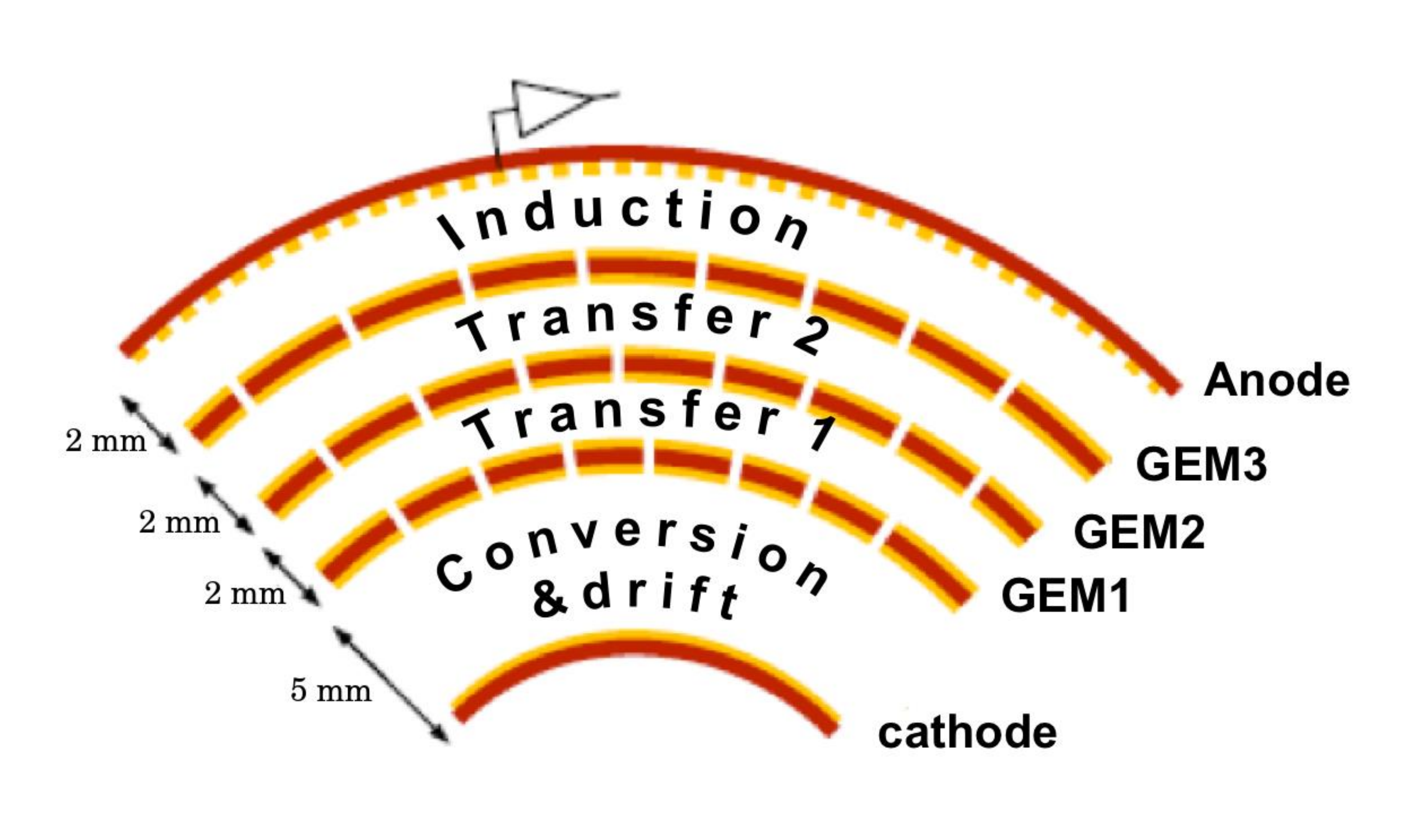}
	\caption{Schematic view of a triple CGEM.}
	\label{fig:cgem}
\end{figure}

In this context, a dedicated front-end readout ASIC delivering the time and charge information for each fired strip has been developed. 

\section{ASIC Architecture}

\IEEEpubidadjcol

\begin{figure}[t]
	\centering
	\includegraphics[width=2.7in]{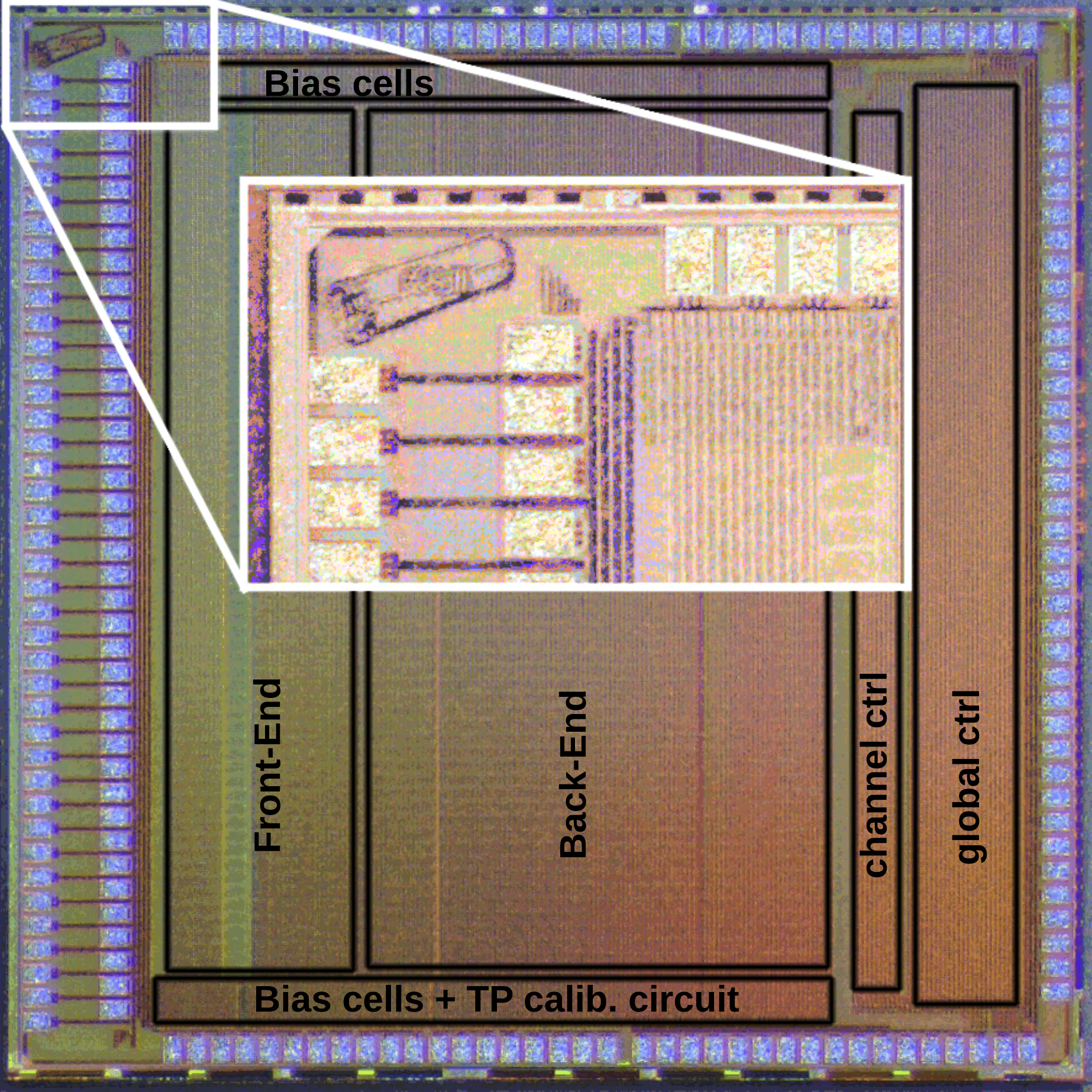}
	\caption{TIGER ASIC.}
	\label{fig:layout}
\end{figure}

TIGER (\textbf{T}urin \textbf{I}ntegrated \textbf{G}em \textbf{E}lectronics for \textbf{R}eadout) is a mixed-signal ASIC for the readout of the CGEM detector. The ASIC architecture is shown in Fig.~\ref{fig:layout} and consists of 64 channels, references and bias generators, an internal test pulse calibration circuitry and a digital global controller. It has been designed in a 110 nm CMOS technology with a die area of 5x5 mm$^{2}$ and a low voltage operation of 1.2 V. The digital back-end is an SEU-upgraded version, i.e. protected against Single Event Upset using Hamming encoding and Triple Modular Redundancy, of the TOFPET2 ASIC~\cite{manuel,agostino}.

The architecture and parameters of the ASIC have been driven by the requirements of the CGEM detector, i.e. 100 pF input capacitance, 60 kHz/channel event rate (including a safety factor of 4). In addition, the charge centroid algorithm requires the ASIC to extract the charge collected by each strip, with a typical input signal that ranges from 3 fC to 50 fC, while a time resolution of about 5 ns is required to time-tag each hit thus enabling the micro-TPC mode. Furthermore, the power dissipation must be lower than 12 mW/channel due to the limited space available in the inner part of the BESIII spectrometer.

\begin{figure}[t]
	\centering
	\includegraphics[width=3.2in]{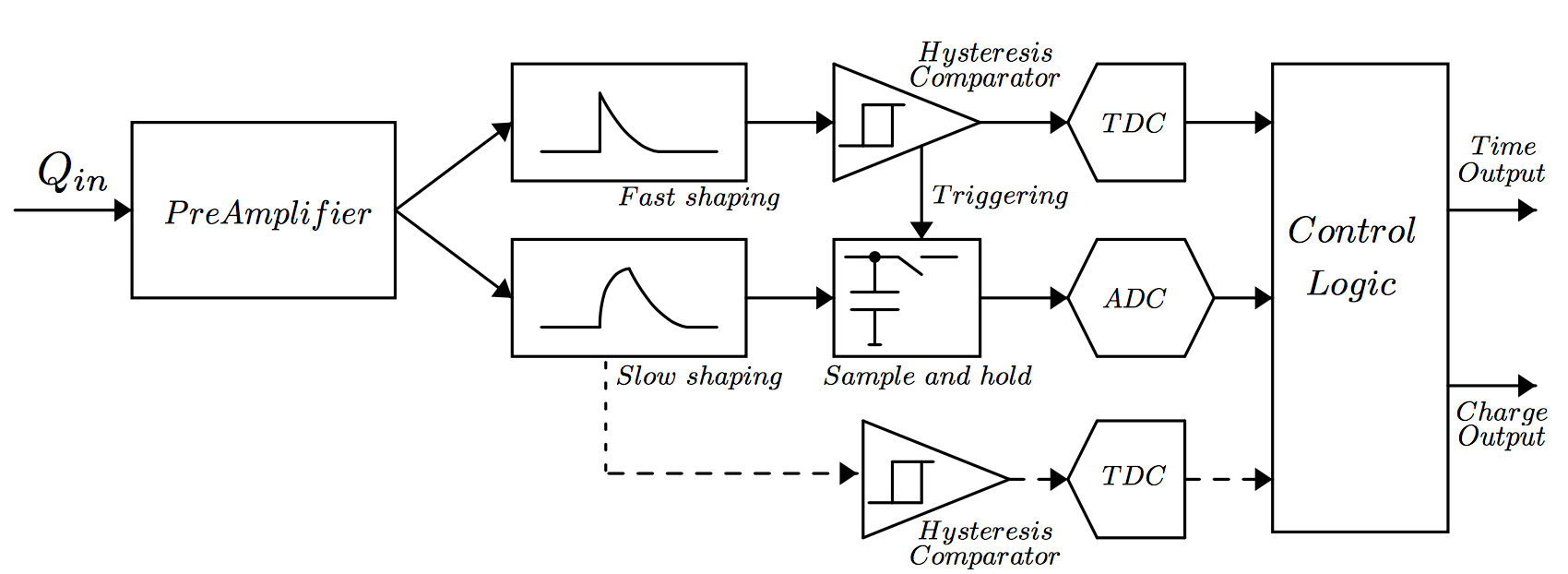}
	\caption{Block diagram of one TIGER channel.}
	\label{fig:schematic}
\end{figure}

A functional block diagram of one channel of TIGER is shown in Fig.~\ref{fig:schematic}. It can be divided into two parts: an analogue front-end to receive the CGEM signals which are amplified and shaped and a mixed-signal back-end for signal digitization, data processing and transmission. Taking into account that the ASIC has to deliver time and energy measurements a dual-branch approach has been adopted. A charge sensitive amplifier generates two replicas of the amplified signal which are sent to two shaping amplifiers: the Time-branch shaper generates a fast signal (60 ns peaking time), optimized for low-jitter timing measurement, while the Energy-branch shaper provides a slower signal (170 ns peaking time), allowing for better signal integration and noise optimization. The outputs of the two shapers are fed to two discriminators with separate and programmable thresholds. 

The channel controller, a digital logic block working at 160 MHz clock frequency, generates the trigger signals for the channel operations. The time and charge information is provided by two low-power TDCs based on analogue interpolation and a sample-and-hold circuit (S/H). Each TDC is composed of a set of four time-to-analogue converters (TAC) and a Wilkinson ADC, and delivers a 50 ps time binning at 160 MHz clock frequency~\cite{rivetti}. The S/H circuit samples the output signal of the E-branch shaper within a time window, generated by the digital logic and configurable by the user. The sampled signal is digitized with the same Wilkinson ADC used for the TDC operation, therefore both TAC and S/H circuits employ a quad-buffer scheme to de-randomize the input event rate and lessen the issue of the inherently high conversion time of this method. The flexibility of this multiple readout scheme allows to extract the charge measurement either from the Time-over-Threshold (ToT) or the S/H circuit digitized value.

The readout scheme features a tigger-less mode in which each input signal above the threshold is digitized and sent to the global controller which manages the data transmission to an FPGA board through 2 LVDS links, using 8B/10B encoding. The ASIC configuration is handled by the FPGA through a 10 MHz SPI like interface and allows to send a test pulse (TP) signal either to the front-end to test the whole readout chain or directly to the channel controller to calibrate the TDCs and assess the functionality of the digital logic of the chip. The analogue TP is generated by an internal calibration circuitry with a user-configurable amplitude, allowing to test the ASIC for the whole input dynamic range.

\section{Test Results}
\begin{figure}[!h]
	\centering
	\includegraphics[width=2.5in]{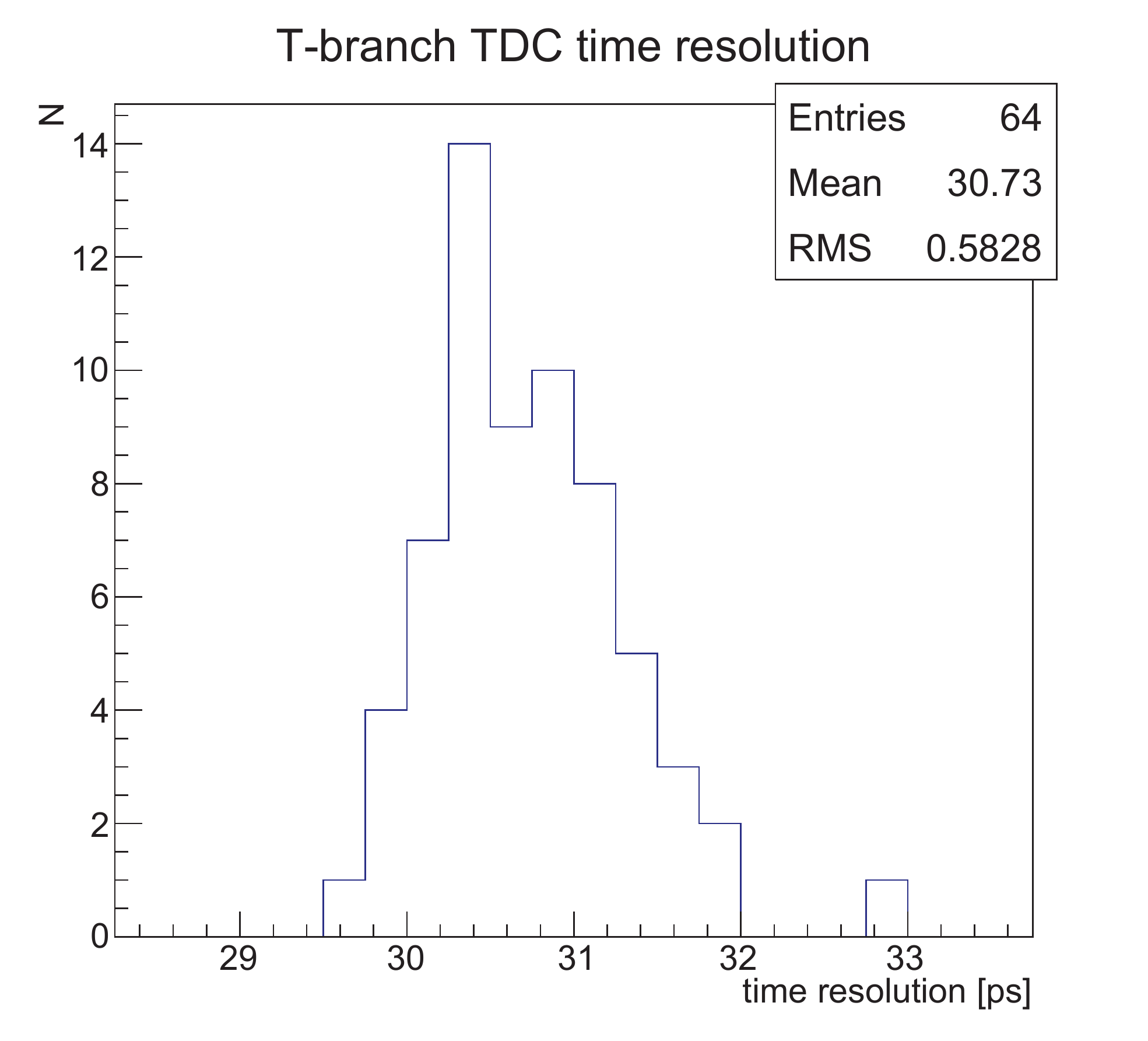}
	\caption{T-branch TDC quantization error.}
	\label{fig:TDC}
\end{figure}

The first prototype was produced in a Multi-Project Wafer (MPW) run and received in October 2016. The ASIC electrical characterization started in November 2016. As a first step, the back-end electronics was tested and found fully functional: Read/Write operations of channel/global configuration registers, data transmission and decoding, TDC operation and fine calibration are working properly.
The TDCs performance have been evaluated by sending a digital test pulse at their inputs and sweeping the TP phase along one clock cycle. A look-up-table (LUT) is generated to store the gain and offset correction for all the channels. After the calibration, the TDC quantization error, shown in Fig.~\ref{fig:TDC}, is lower than 50 ps r.m.s for both branches and for all the channels. 

This implies that the system time resolution is affected only by the sensor and the analogue front-end response. The jitter of the T-branch has been evaluated using the internal calibration circuitry to generate signals of fixed charge (3 fC) which are fed at the preamplifier input. The measured time jitter is below 5 ns r.m.s., allowing for efficient time-tag.

\vspace{1cm}

\begin{figure}[t]
	\centering
	\includegraphics[width=2.8in]{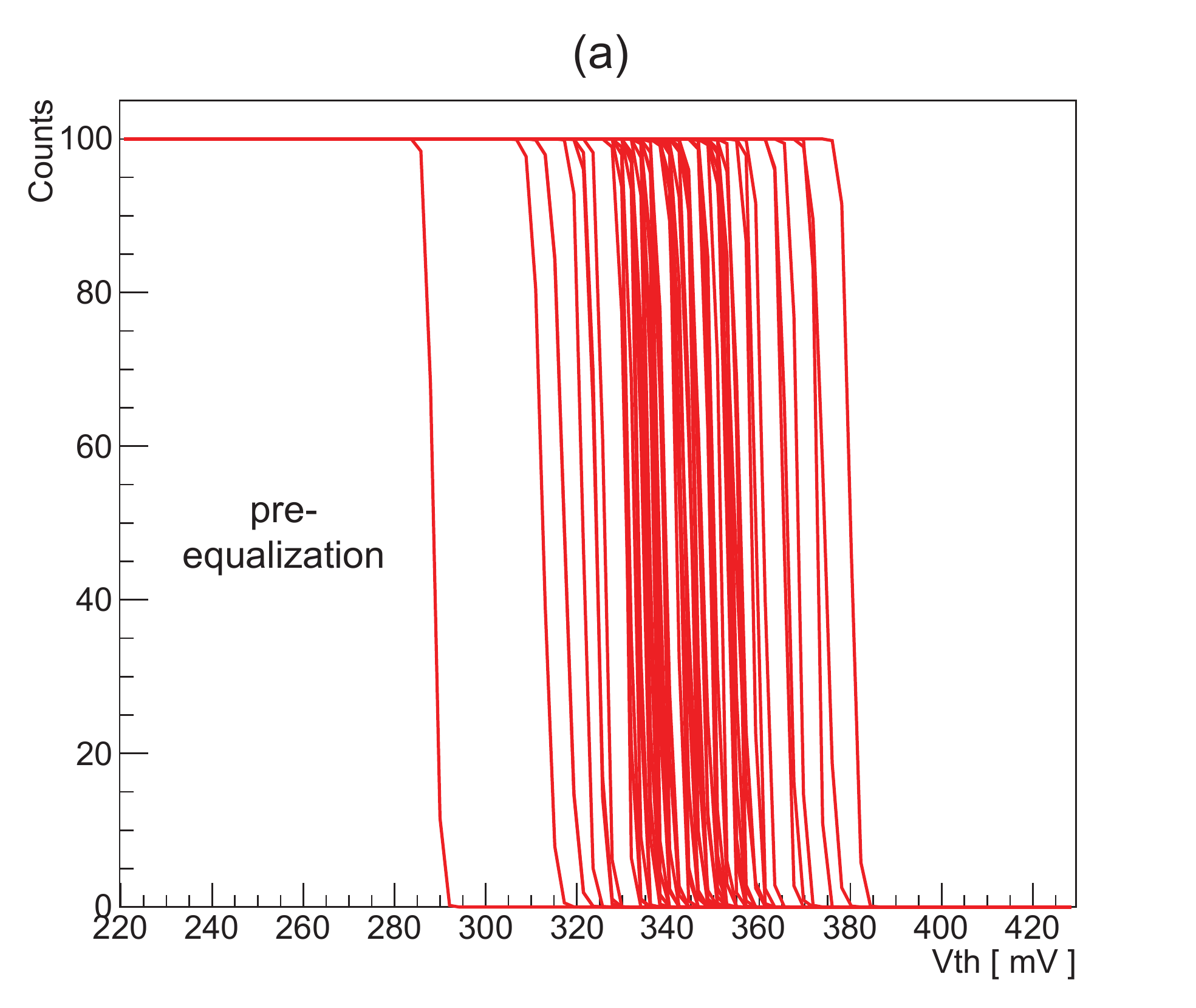}
	\centering
	\includegraphics[width=2.8in]{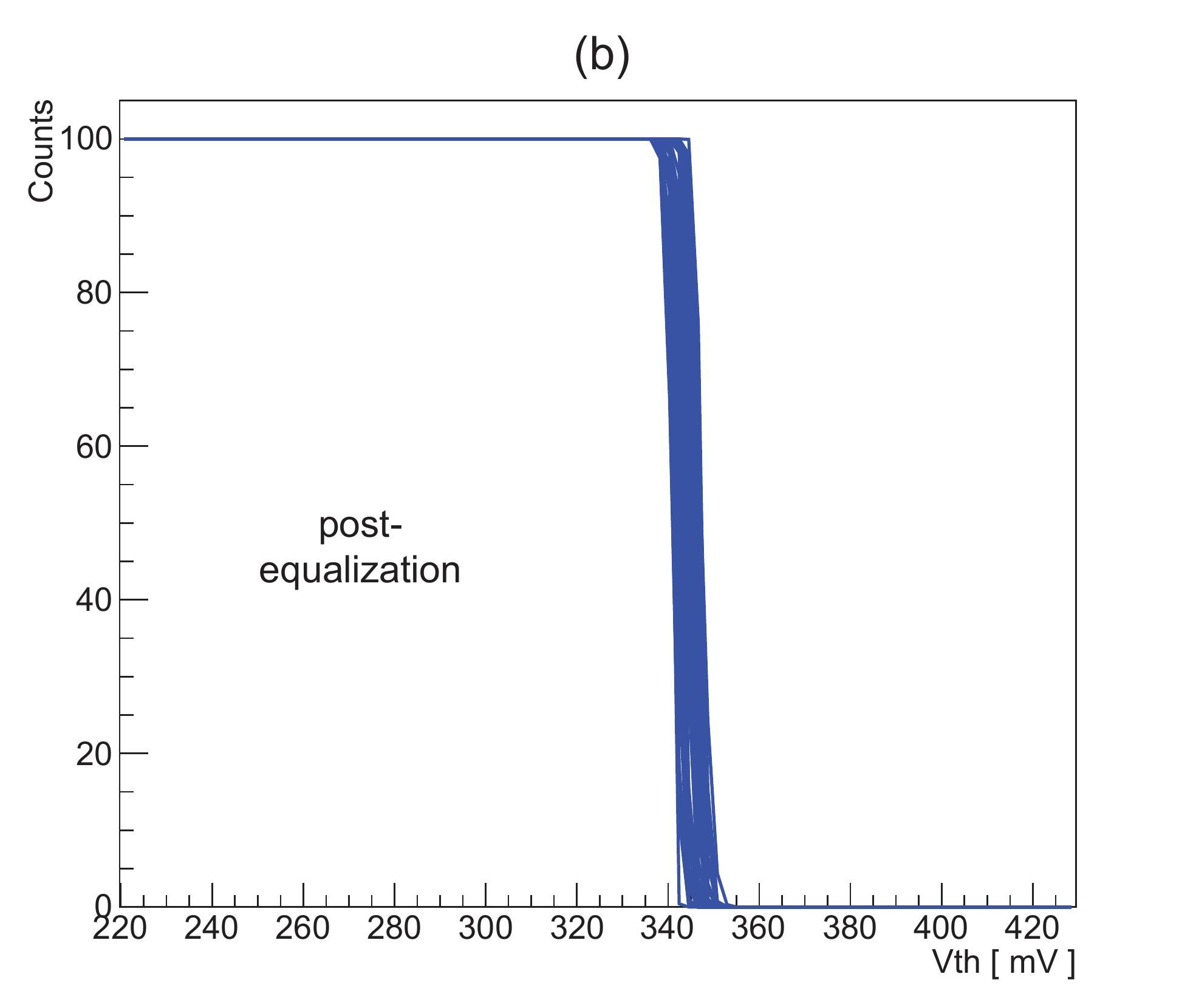}
	\caption{S-curves for the 64 channels before (a) and after (b) thresholds equalization.}
	\label{fig:VTH}
\end{figure}

Threshold scans have been performed for thresholds equalization and noise measurement.
In this method, the s-curve produced by the threshold scan is fitted with a sigmoid function in which the 50\% value corresponds to the baseline level and the sigma provides the noise. A look-up-table allows to correct the mismatches between the channels by adjusting their effective threshold. Fig.~\ref{fig:VTH} shows the s-curves for the 64 channels before and after thresholds equalization.

\begin{figure}[t]
	\centering
	\includegraphics[width=3.5in]{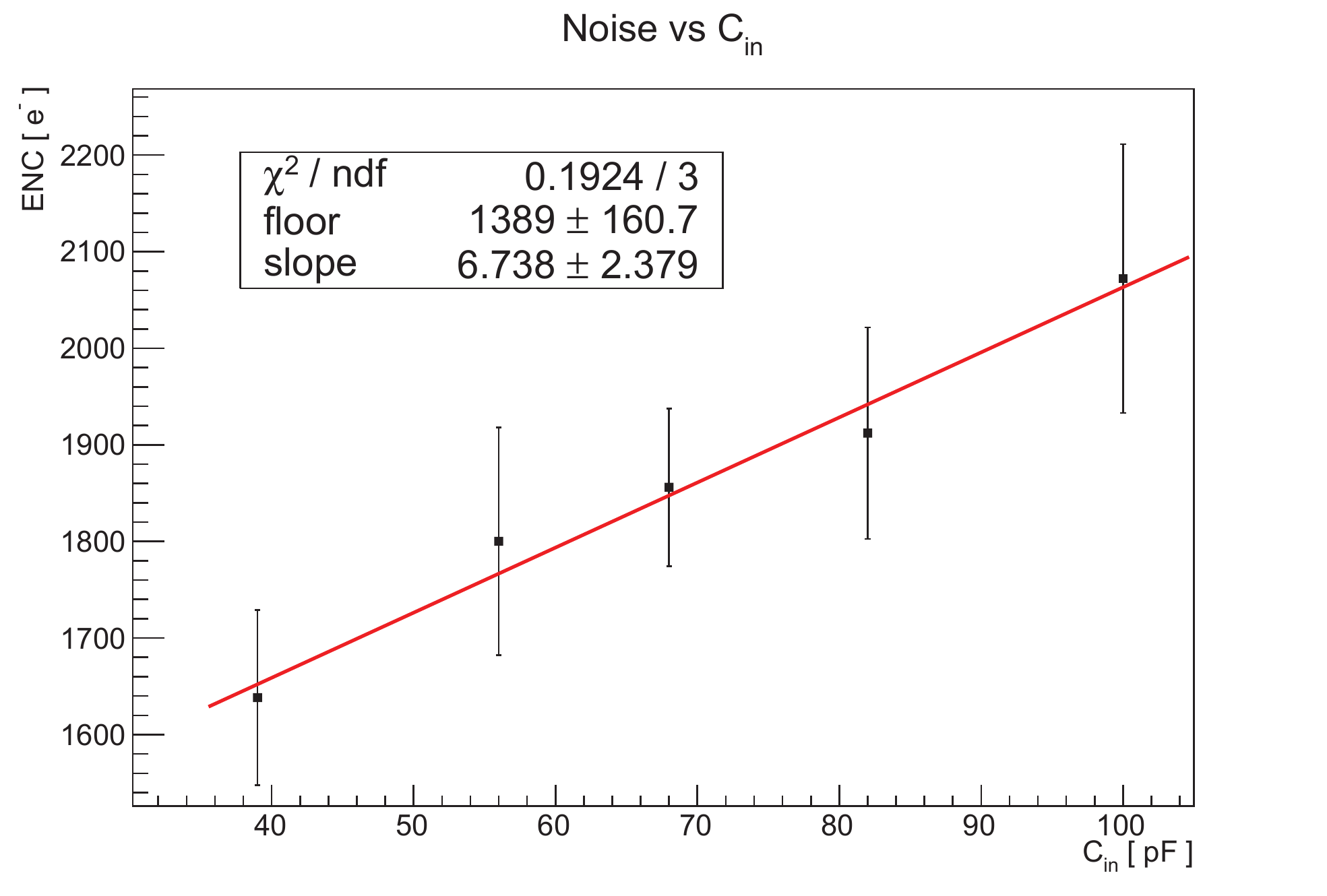}
	\caption{Equivalent Noise Charge.}
	\label{fig:ENC}
\end{figure}

The noise has been evaluated for different input capacitances and the results are displayed in Fig.~\ref{fig:ENC}. The equivalent noise charge (ENC) measured at the output of the E-branch shaper for a 100 pF input capacitance is about 2100 electrons, which is slightly higher than the value expected from simulations. Despite the fact that the measured performance are already adequate for our application, more dedicated test are currently ongoing in order to reduce the noise.

\section{Outlook and Conclusions}
A 64-channel ASIC, named TIGER, has been developed to readout the signals coming from the BESIII CGEM detector. The results from the first prototype electrical characterization are in good agreement with the expected values, except for the noise, and the performance meet the requirements set by the experiment. Tests with the detector are currently ongoing using cosmic rays and radioactive sources, in order to fully qualify the ASIC with the sensor. A new version has been submitted in August 2017, with minor design revisions, to fully equip the detector for the  installation which is scheduled for Summer 2018.

% if have a single appendix:
%\appendix[Proof of the Zonklar Equations]
% or
%\appendix  % for no appendix heading
% do not use \section anymore after \appendix, only \section*
% is possibly needed

% use appendices with more than one appendix
% then use \section to start each appendix
% you must declare a \section before using any
% \subsection or using \label (\appendices by itself
% starts a section numbered zero.)
%

%\appendices
%\section{Proof of the First Zonklar Equation}
%Appendix one text goes here.

% you can choose not to have a title for an appendix
% if you want by leaving the argument blank
%\section{}
%Appendix two text goes here.

\section*{Acknowledgment}
This research activity has been performed within the \mbox{BESIIICGEM} Project, funded by European Commission in the call H2020-MSCA-RISE-2014.

% Can use something like this to put references on a page
% by themselves when using endfloat and the captionsoff option.
\ifCLASSOPTIONcaptionsoff
  \newpage
\fi

\end{document}